      [1999/12/01 v1.4c Il Nuovo Cimento]
 \documentclass{cimento}

\usepackage{graphicx}  

\title{Analysis of X-ray flares in GRBs}

\author{D.~Guetta\from{ins:api}\ETC\thanks{guetta@mporzio.astro.it},
V.~D'Elia\from{ins:api},
F.~Fiore\from{ins:api},
M.L.~Conciatore\from{ins:leic},
A.~Antonelli\from{ins:api}
L.~Stella\from{ins:api}}

\instlist{
\inst{ins:api}INAF --- Osservatorio Astronomico di Roma, via Frascati 33,
Monteporzio Catone (Roma), Italy
      \inst{ins:leic} ASDC
}

\PACSes{
\PACSit{98.70.Rz}{$\gamma$-ray sources; $\gamma$-ray bursts}
\PACSit{95.85.Nv}{X-ray flares}
}

\begin{document}

\maketitle

\begin{abstract}
 We present a detailed study of the spectral and temporal
properties of the X-ray flares emission of several GRBs. 
We select a sample of GRBs which  X-ray light curve exhibits large
amplitude variations with several rebrightenings superposed on the
underlying three-segment broken powerlaw that is often seen in
Swift GRBs. We try to understand the origin of these
fluctuations giving some diagnostic in order to discriminate 
between refreshed shocks and late internal shocks.
For some bursts our time-resolved spectral analysis supports the
interpretation of a long-lived central engine, with rebrightenings
consistent with energy injection in refreshed shocks as slower
shells generated in the central engine prompt phase catch up with
the afterglow shock at later times

\end{abstract}


\section{Introduction}

Most of the Gamma Ray Bursts (GRBs) observed  in the pre-Swift 
era showed a smooth power law decay with time in the X-ray and
optical afterglow light-curve \cite{ref:laurs03}. 
This behavior was found to be consistent with the basic predictions of the
fireball model, where the afterglow flux is produced as a relativistic blast
wave propagating into an external medium. Under the assumption of a
spherical fireball and a uniform medium, Sari, Piran \& Narayan (1998)
\cite{ref:sari98} showed that the dependence of the flux on frequency and time can be
represented by several power law segments, $F_{\nu}\propto
\nu^{-\beta}t^{\alpha}$, each of which applies to a different regime. 
Before Swift only a few GRBs showing
deviations from the smooth power law light curve were known, see e.g.
the case of GRB 021004 \cite{ref:bers02},\cite{ref:math02}.

This simple picture is now changing since the advent of Swift.
Erratic X-ray flares have been detected by Swift in almost
half of its GRBs (\cite{ref:geh05}, \cite{ref:bur05}, 
\cite{ref:falc06}, \cite{ref:nous06}, \cite{ref:obr06}). 
While some bursts show one distinct flare after the GRB prompt emission,
(e.g. GRB 050406 \cite{ref:rom06}), 
other events like GRB 050502B and GRB 050713A,
show several flares \cite{ref:bur05}, \cite{ref:obr06}, \cite{ref:falc06}.
One of the main  goals of current GRB
science is to understand the origin of this newly observed light
curve behavior. Since flares are likely to trace the activity of
the central engine, they can help us gain insights on
the physical processes governing the early post-GRB phases. 
Note that flares have been observed in the light curves of both long
and short GRBs.  As a first step toward the understanding of
different flares characteristics it is fundamental to carry out
a systematic investigation of the morphological and timing properties
of the flares observed by Swift/XRT (Chincarini et al. 2006\cite{ref:chinca06}).

In some cases we can already rule out some models that have been
proposed to explain the flare phenomenon.
The presence of several flares in one GRB strongly disfavors scenarios that
can only account for one flare, e.g. the self synchrotron Compton 
emission from reverse shock
\cite{ref:koba06}. Density bumps \cite{ref:laz02} cannot be responsible of 
large amplitude rebrightenings (e.g. GRB 050502B \cite{ref:falc06}) .
In this paper we carry out  a detailed
spectral and timing analysis of flares in several GRBs with the goal of
constraining the physical mechanism responsible for its
production. In particular we concentrate
on two models proposed for the flares, the late internal shock (IS) model
and the refreshed shocks (RS) model.

In the internal shock scenario the flares are due to the reactivation
of the GRB central engine and they are produced from an ``internal'' dissipation
radius within the external blast wave front \cite{ref:zhang06}.
In this case a mechanism to reactivate the engine is required.
For the long bursts, King et al. (2005) \cite{ref:king06} proposed a
model in which the flares could be produced from the fragmentation
of the collapsing stellar core in a modified hypernova scenario. For
the case of short GRBs, MacFadyen et al. (2005) \cite{ref:mac05}
suggested that the
flares could be the result of the interaction between the GRB
outflow and a non-stellar companion. More recently, Perna et al. (2006)
\cite{ref:perna06} have analyzed the observational properties of flares in both
long and short bursts, and suggested a common scenario in which
flares are powered by the late-time accretion of fragments of
material produced in the gravitationally unstable outer parts of an
hyper-accreting accretion disk.  Other mechanisms that could produce
flares involve a magnetic origin \cite{ref:gao06}, \cite{ref:proga06}.

In the refreshed shock scenario \cite{ref:kum00} the flares are due to
late time energy injections into the main afterglow 
shock by slow moving shells ejected from the central engine during the prompt phase.
In this case we do not need a reactivation of the engine.
This model can interpret several flares features \cite{ref:guetta06},
\cite{ref:perri06} seen in some GRBs.

Both models have observational consequences that can be tested.
In this paper we propose and exploit some diagnostics between the two models.

\section{The sample and the data analysis}

We consider all the GRBs detected by Swift between the Swift launch and
2006 January 31. We select  all those that showed 
prominent flares in their X-ray lightcurves. 
In table 1 we list all the GRBs that were selected for the analysis. 
A more extensive analysis that will be presented includes a larger sample 
of Swift GRBs \cite{ref:guetta06b}.
The X-ray data were reduced using the Swift Software (v. 2.0) and in
particular the XRT software developed at the ASDC and HEASARC
\cite{ref:capa05} \footnote{http://heasarc.gsfc.nasa.gov/docs/swift/analysis/xrt\_swguide\_v1\_2.pdf}.

Important information on the mechanism producing the flares can be
obtained from the comparison between the observed variability timescale
$\Delta T$ and the time $T$, at which the flare is observed. 
In the refreshed shock scenario the early afterglow has
a variability timescale of the order of the time since the explosion,
$\Delta T\sim T$ \cite{ref:kum00}, while in the late 
internal shock model $\Delta T << T$.
We calculated for all the flares of the sample the ratio between the duration
of the flare, $\Delta T$, and the time at which the flare peaks $T_{\rm peak}$
and report these values in Table 1. We also report the time at which 
the rising behavior of the light curve begins $T_{\rm rise}$.
In our analysis $\Delta T$ is defined as the difference between the times 
at which the light
curve shows the end of the falling and the beginning of the rising
behaviors. All  times  refer to the BAT trigger.
As we can see from this table, within the uncertainties, all flares
have $\Delta T/ T_{\rm peak}\sim O(1)$, in agreement with the refreshed
shock scenario. However some of these flares have other characteristics
that can be better explained in the internal shock scenario.

The strategy for our temporal fitting procedure is the following. We
consider the data in the range between $T_{\rm rise}$ and 
$T_{\rm peak}$ to compute the
slope,$\alpha_r$, of the rising part of the light curve. 
The start time is set to $T_{\rm rise}$. 
For the decay, we use the data in the range between $T_{\rm peak}$ and the
end of the decaying behavior to estimate the power law decay index
 $\alpha_d$. The start time has been set to $T_{\rm peak}$.
We  note that the power-law 
rise and decay indices depend strongly on the initial counting
time, $T_0$.
In the internal shock scenario, the re-starting of the central engine
re-sets the starting time. This is why some authors \cite{ref:liang06}
use $T_0\sim T_{\rm rise}$ to estimate $\alpha_d$ and thus find
 much steeper temporal slopes (we call these slopes $\alpha_{\rm d,IS}$). 
We think that the right choice for the start time is $T_{\rm peak}$. 
This is
because a start time earlier than the beginning of the decaying behavior
causes an unphysical steepening of the slope, due only to mathematical
effects.
In any case, since the $T_0$ issue is an open problem,
we also discuss what happens to the decay 
slope if the start time is set to
$T_{\rm rise}$. We therefore report both $\alpha_{d}$ and $\alpha_{\rm d,IS}$
for the bursts were the internal shocks better interpret the flare
phenomenology.

\subsection{Time-resolved spectroscopy}

The analysis of the window timing (WT) light curves of the 
flares reveal complex spectral variations. 
To investigate  the nature of these variations
we performed a time resolved spectral analysis.
We first fitted the spectra in the rise and decay parts 
with a simple power law combined with
photoelectric absorption. We indicate with $\Gamma=\beta+1$ the
photon index. The results of our fits are shown in Table 2.  Large
spectral variations are evident, in particular between the spectra
corresponding to the rise and  fall of several flares. 

Motivated by the synchrotron emission model \cite{ref:sari98}, 
we then fitted the spectra using a broken power law
model.  We kept the low energy (before the cooling frequency)
spectral power law index fixed at
$\beta_1=\Gamma_1 -1=0.5$, as expected from standard synchrotron
models and left the high energy power law index
$\beta_2=\Gamma_2-1$ as a free parameter. The results are shown in
Table 2. In some cases we were not able to fit the spectrum with a 
broken power law, as evident in Table 3 (we write ``na'' in this case).

\begin{table}[ht!]
\caption{\bf values of z, $\Delta T$, $T$ for each flare}
\begin{tabular}{lccccc}
\hline
\hline
GRB         & z      & $T_{\rm rise}$ & $T_{\rm peak}$ &$\Delta T$ &$\Delta
T/T_{\rm peak}$  \\
\hline
050406      &        & 132      & 212      & 160       & 0.75 \\
050502B     &        & 468      & 848      & 700       & 0.83 \\
050713A     &        & 97       & 107      & 50        & 0.47 \\
050730(1)   & 3.97   & 202      & 232      & 100       & 0.43 \\
050730(2)   & 3.97   & 302      & 432      & 300       & 0.69 \\
050730(3)   & 3.97   & 602      & 672      & 340       & 0.51 \\
050822      &        & 211      & 231      & 70        & 0.30 \\
050904      & 6.29   & 368      & 458      & 200       & 0.44 \\
050908      &        & 268      & 388      & 500       & 1.29 \\
060111A     &        & 204      & 274      & 260       & 0.95 \\
060124(1)   &        & 311      & 571      & 330       & 0.58 \\
060124(2)   &        & 641      & 701      & 230       & 0.33 \\
\hline
\end{tabular}
\end{table}

\section{Refreshed or late internal shocks?}
In this section we try to distinguish
between the two (IS and RS) models using the temporal and spectral
properties derived for each flare in the previous section.
As we have emphasized, the parameter that helps in this analysis is 
the flare duration, $\Delta T$.

\subsection{Late-time ``energy injection''  }

If {\bf $\Delta T\sim T$ }
rebrightenings and bumps in the afterglow light curve may be due 
to ``energy injections'' at later times that could be
produced by slower shells that catch up with the afterglow shock at
later times. In this case we have some predictions on how the temporal
and the spectral energy slope should be related.
We consider the standard afterglow model in the adiabatic evolution 
\cite{ref:sari98}. The analysis can be easily extended 
to the radiative evolution and we will do this in our future paper 
\cite{ref:guetta06b}.
We can distinguish four main cases that we show in Fig. 1:
\begin{itemize}
\item (i) The spectral break  frequencies $\nu_b$,
(where we have the change in the spectral slope) are 
within the 0.5--10 keV band covered by the XRT
observation, both
during the rise ($\nu_{b,r}$) and the decay
($\nu_{b,d}$) phases. 
In this case $\alpha_d = -1/4$ for $\nu<\nu_{b,d}$ and 
$\alpha_d = -3p/4+1/2$ for $\nu>\nu_{b,d}$.
Since the spectral index in the decay, $\beta_d$, is expected to be 
$\beta_d=-p/2$ after the break (as expected in the synchrotron model),
a relation between the spectral and
temporal slope as, $\alpha_d =3/2\beta_d+1/2$, is expected.  
 Indeed, for a completely adiabatic
evolution, one has $\nu_b\propto E^{1/2}$ (where E is the burst
energy), while for a fully radiative evolution $\nu_b \propto
E^{4/7}$, both implying that the flux rebrightening $F_2/F_1$ scales
with  the frequency change factor as
$\left(\nu_{b,r}/\nu_{b,d}\right)^2$.
As an example of this class we consider the GRB 050713A 
\cite{ref:guetta06} and
in particular the first flare detected by XRT. In this case
$\alpha_d$ follows the model prediction as shown in Fig.2; 
it goes from $\alpha_d=-0.25$ for $\nu<\nu_{b,d}$, to $\alpha_d=-1$
for $\nu>\nu_{b,d}$.
The predicted rebrightening is
$F_2/F_1 \sim \left(\nu_{b,r}/\nu_{b,d}\right)^2\sim 10$,
similar to the observed value.

\item (ii) The synchrotron frequency $\nu_{b,r}$ 
is outside of the band during the rise $\nu_{b,r}>10$ keV, 
as implied by the single
power-law  fit to the data and  $\nu_{b,d}$ 
is within the 0.5--10 keV band covered by the XRT
observation.
In this case $\alpha_d = -1/4$ for $\nu<\nu_{b,d}$ and 
$\alpha_d = 3/2\beta_d+1/2$ for $\nu>\nu_{b,d}$ and 
$F_2/F_1 \geq \left(10/\nu_{b,d}\right)^2$.

\item (iii) The synchrotron frequencies $\nu_{b,r}$
is outside of the band during the rise  $\nu_{b,r}>10$ keV, 
and  shifts below the lower limit of the energy band  $\nu_{b,d}<0.5$ keV
during the decay. In this case $\alpha_d = 3/2\beta_d+1/2$ 
and $F_2/F_1 \geq \left(10/0.5\right)^2$.

\item (iv) The synchrotron frequencies $\nu_{b,r}$
is outside of the band during the rise and decay phase
 $\nu_{b,r}<0.5$ keV. 
In this case $\alpha_d = 3/2\beta_d+1/2$, but we do not have any constraint on
$F_2/F_1 $. As an example of this class we consider GRB 050730
 \cite{ref:perri06} and in particular its second flare shown in Fig. 3. 

From the best fit spectral energy index during the decay,
$\beta_{\rm d} = 0.70\pm 0.07$, we find $\alpha_d=-0.55$
which is consistent with the temporal decay index found from the temporal fit, 
$\alpha_d=-0.67\pm 0.16$ as we can see in Table 2.
Also the flare of GRB050904 satisfies these relations. 

\end{itemize}

\begin{figure}
\begin{tabular}{llcc}
\includegraphics[angle=-90,width=8truecm]{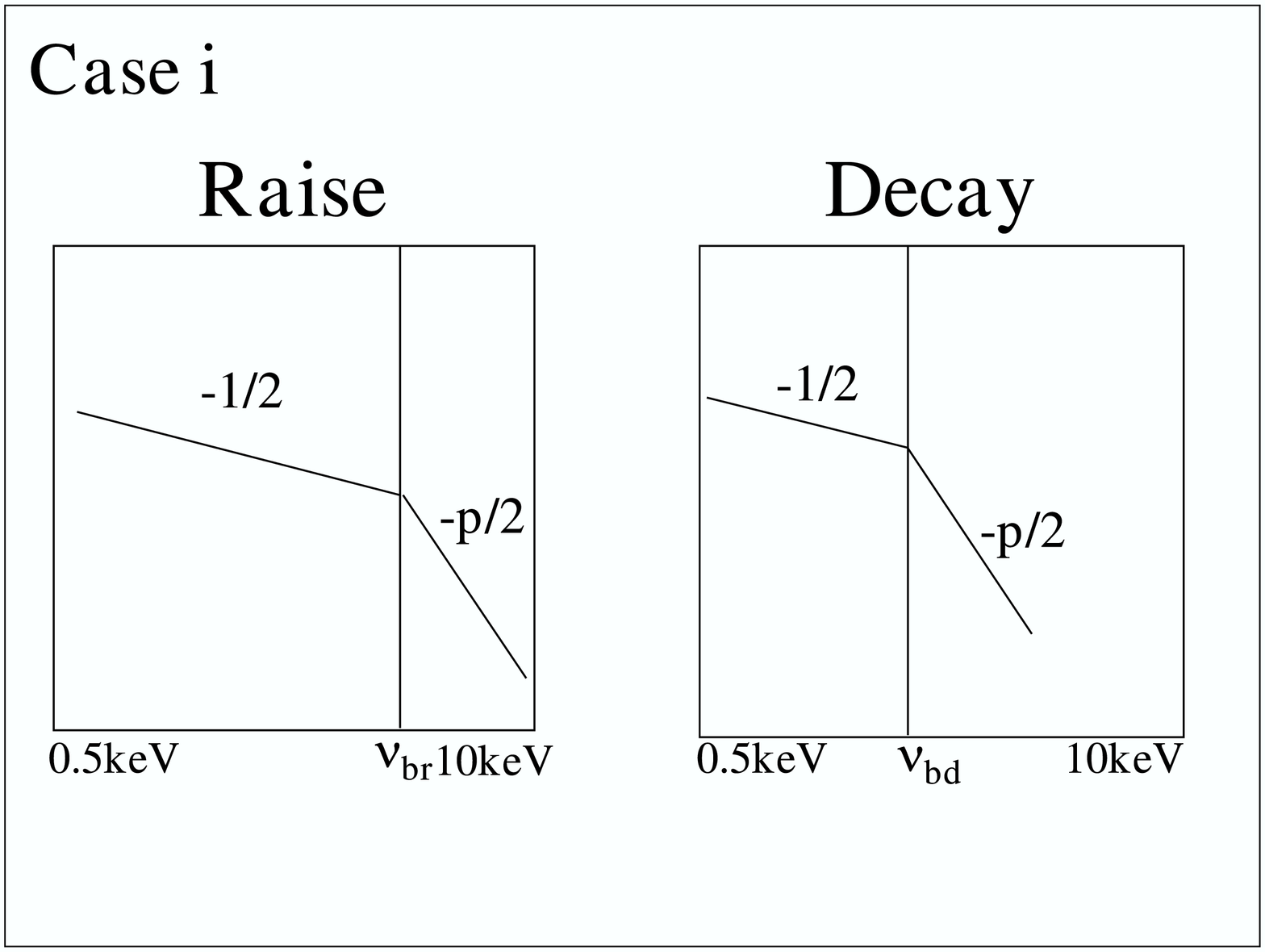}
\includegraphics[angle=-90,width=8truecm]{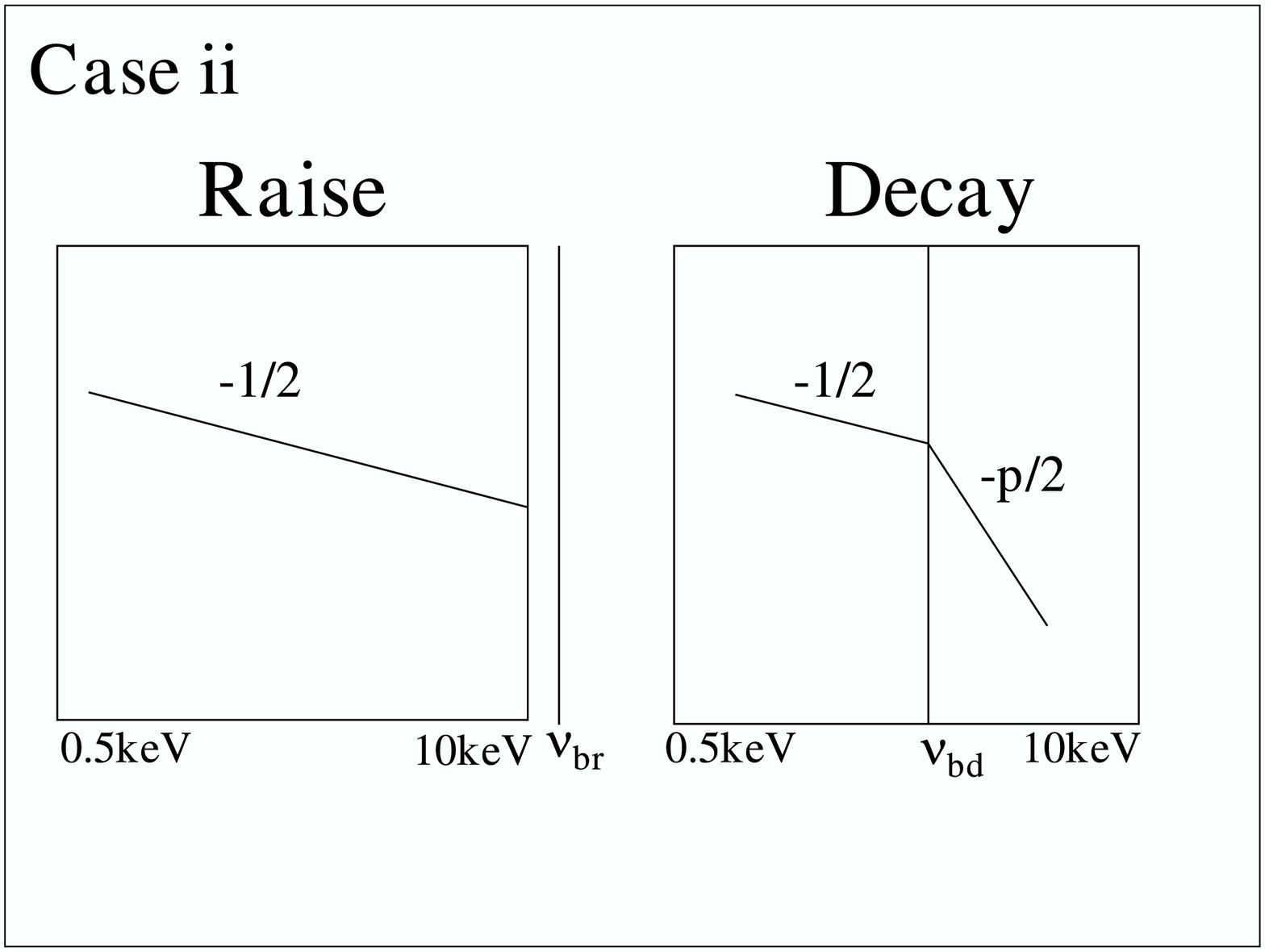}\\
\includegraphics[angle=-90,width=8truecm]{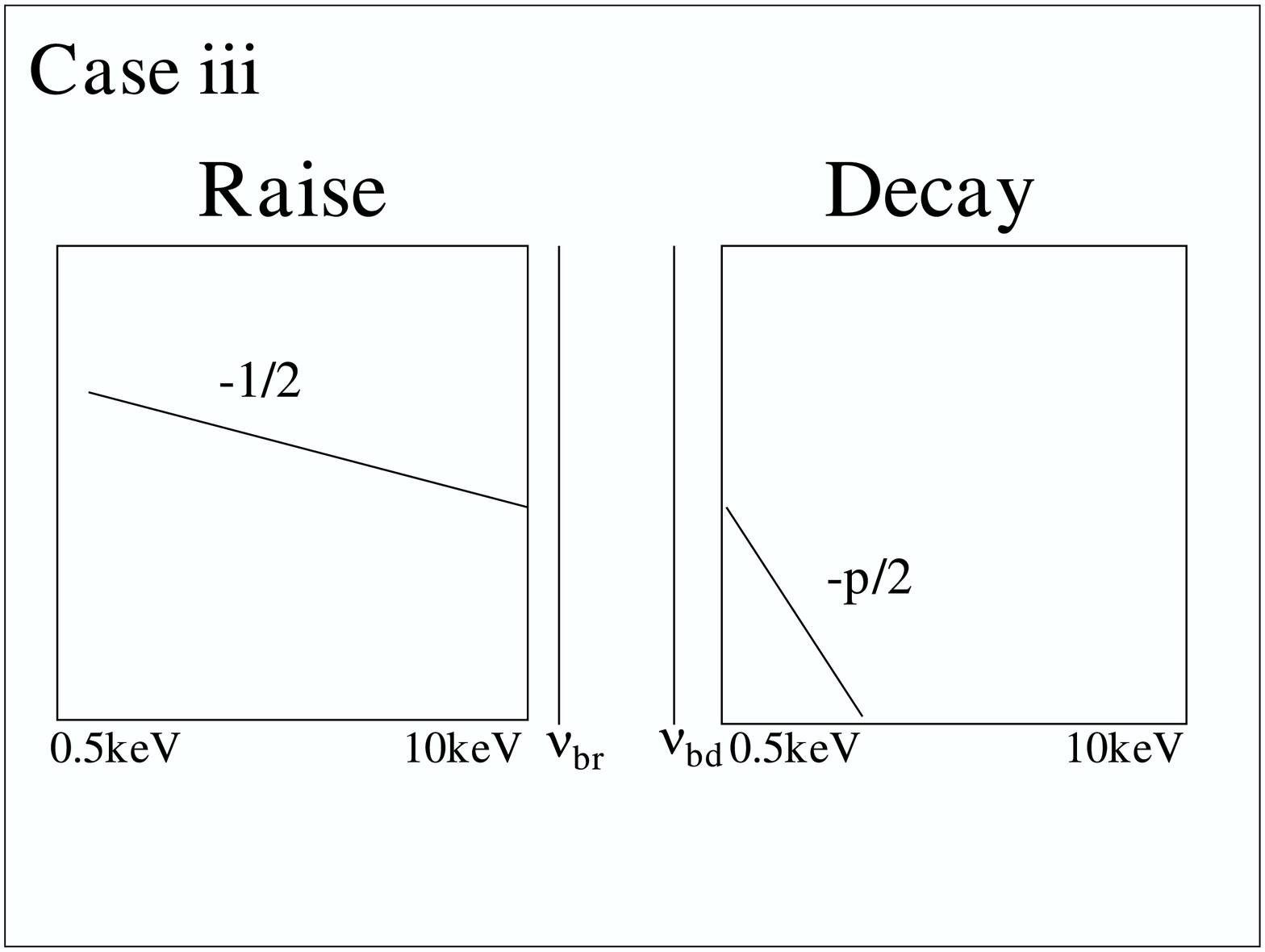}
\includegraphics[angle=-90,width=8truecm]{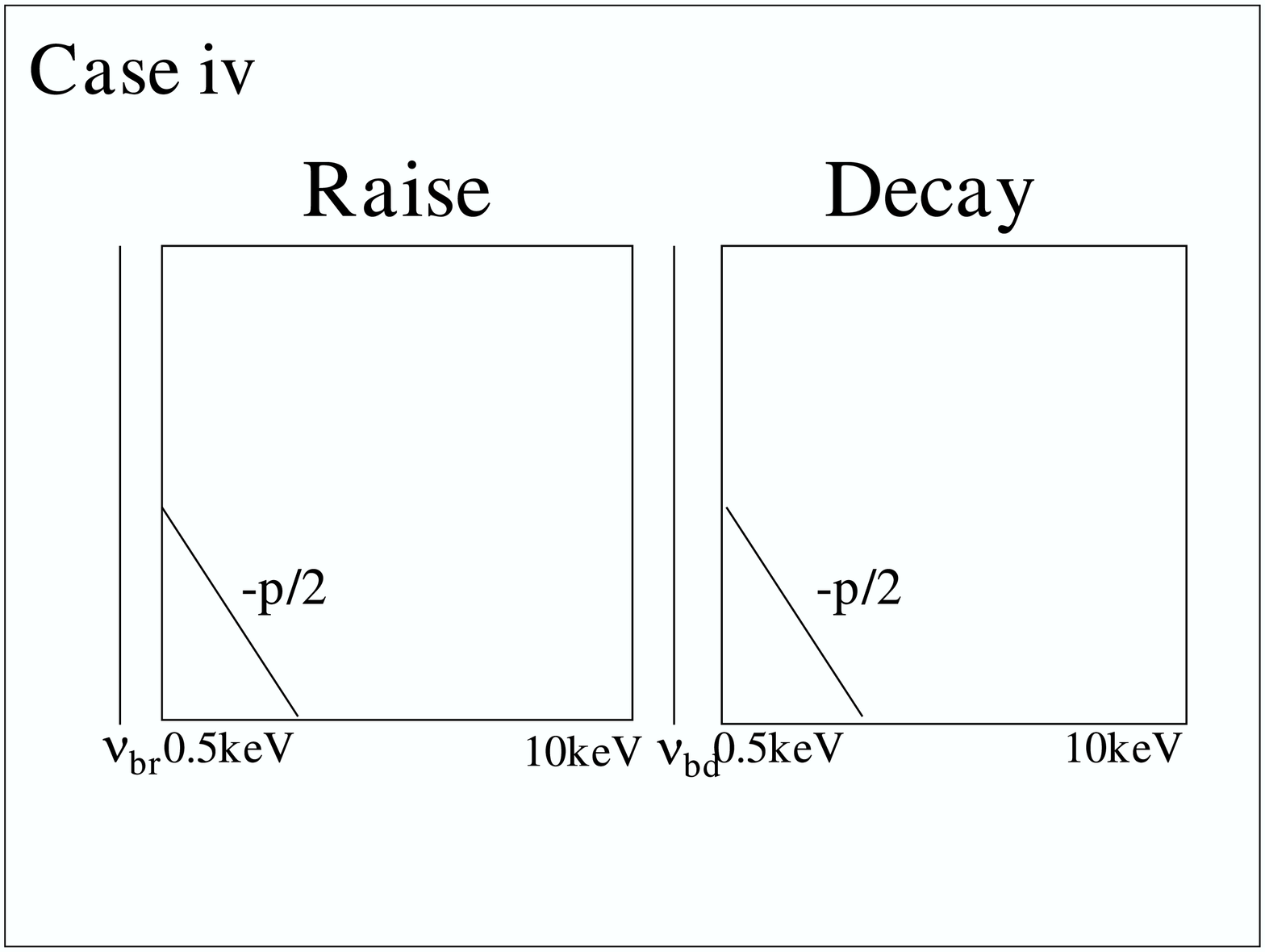}
\end{tabular}
\caption{The four cases in the refreshed shock model explained in sec.~3.1}
\label{cases}
\end{figure}

\begin{figure}
\centering
\includegraphics[width=8cm,height=8cm]{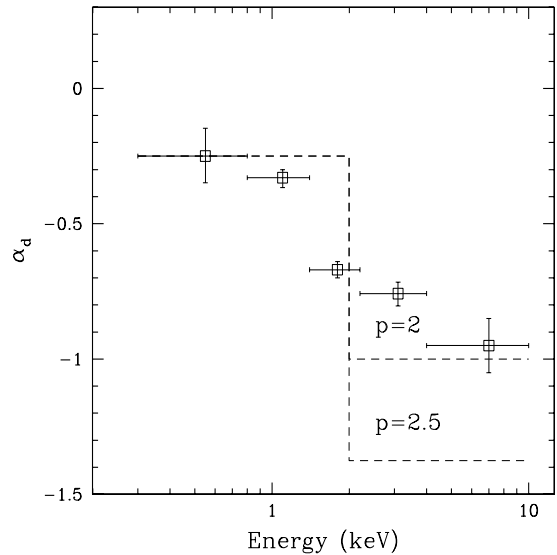}
\caption{\label{discesa} Power-law decay indices of the GRB050713A second
  flare light curves
in the 0.3--0.8, 0.8--1.4, 1.4--2.2, 2.2--4, and 4--10 keV bands,
starting from 108~s after $t_0$, as a function of the energy. The
prediction of the synchrotron model is represented by the dashed lines,
considering a broken power-law  for two different values of $p$.}
\end{figure}

\begin{figure} 
  \centering 
  \includegraphics[angle=-90,width=95mm]{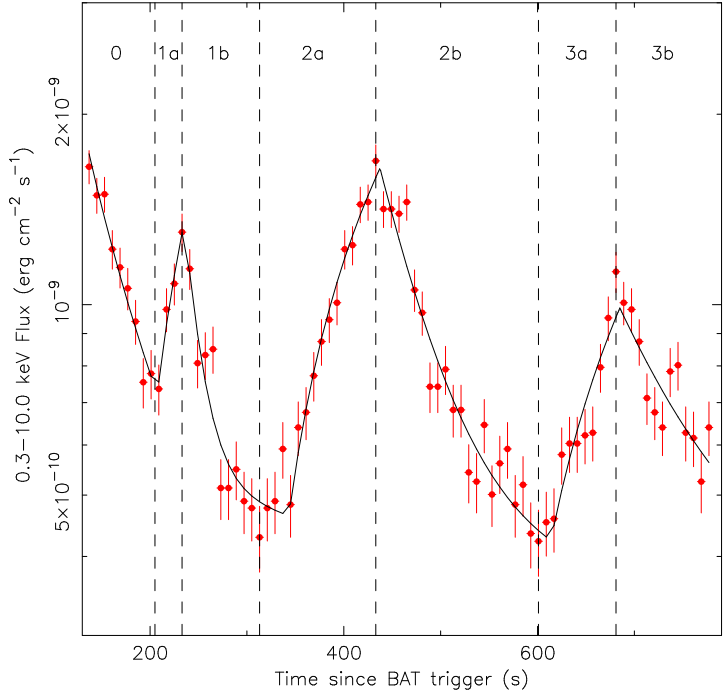} 
  \caption{{\it Swift} XRT 0.3-10 keV light curve of the GRB 050730 
X-ray afterglow during the first orbit. The solid line is the best fit model
to the data obtained considering a linear rise exponential decay for the three
flares (see Sect.~\ref{temporal} of Perri et al. 2006).
               The dashed vertical lines delimit the seven time intervals 
               considered for the spectral analysis. Data are binned to 8 seconds resolution and errors are 
               at the 1$\sigma$ level.} 
  \label{lc} 
\end{figure}

\subsection{Late internal shocks}
({\bf $\Delta T<< T$ })

A diagnostic to check if the re-brightenings are due to late 
internal shocks has been recently proposed by Liang et al. 2006
\cite{ref:liang06}. 
In the internal-origin scenario for X--ray flares the decay emission of 
the flaring episodes should be dominated by high-latitude emission with 
the temporal decay index related to the decay spectral index 
as $\alpha_{\rm d}=\beta_{\rm d}+2$ (\cite{ref:kumar00b}).
Liang et al.~2006 use the time integrated spectrum;
however in order to test the curvature effect,
the analysis should be done using the 
spectrum in the decay phase of the flare.
We have checked this interpretation for the first three bright X--ray 
flares observed in  GRB050730 afterglow shown in Fig.3 
and we plan to extend the analysis to all  flares in our sample
\cite{ref:guetta06b}. 
We used the spectral indices measured during the decay  
and  found best fit zero times \cite{ref:perri06} 
$t_{0,1}=77^{+22}_{-29}$ s, $t_{0,2}=234^{+52}_{-85}$ s 
and $t_{0,3}=174^{+99}_{-152}$ s, respectively. 
Comparing these values with the rise times given in Table 1,
we can see that for all three episodes the zero time values are not 
located during the rising portion 
of the corresponding flare, clearly indicating that the observed decay 
slopes are not consistent with being due to 
high-latitude emission, as predicted by the late internal shock scenario.

Nonetheless there are flares that can be better explained with the internal
shock model even if $\Delta T\sim T$. These are for example  
GRB050406 and GRB050502B where the presence of an underlying 
continuum in the X-ray lightcurve, consistent 
with the same slope before and after the flare argues against
the external shock models in the adiabatic regime,
 since no trace of an energy injection can be found. 
However it is possible that for these flares 
 the radiative regime applies \cite{ref:guetta06b}.
If we consider $T_0=T_{\rm rise}$ we estimate 
$\alpha_{\rm d,IS}=-3.2\pm 0.7$, $\alpha_{\rm d,IS}=-4.3\pm 0.1$
and $\alpha_{\rm d,IS}=-1.4\pm 0.2$ for GRB050406, GRB050502B and GRB050822
(that is the only one that has $\Delta T<<T$).
We can compare these values with the ones expected by the curvature effect.
For GRB050406 we find $\alpha_{\rm d,IS}=-(\beta_d+2)=-3.2$, in agreement with
the observed value, whereas for GRB050502B and GRB050822
we find $\alpha_{\rm d,IS}=-3.3$ and $\alpha_d=-3.8$, which are different 
from the observed values.

\begin{table}[ht!]
\caption{\bf Best fit values for the temporal and spectral power slopes. 
We also indicate which models best interprets 
the flares phenomenology,  internal shocks (IS) and
refreshed shocks (RS). For the RS model we also report to which case
of the ones discussed in sec~3.1 the GRB belongs}
\begin{tabular}{lc|cccc}
\hline
\hline
GRB     &$ \alpha $  & N$_H$              & $\Gamma$      &   $\chi^2_{\rm R}$
& model  \\
           &  & $10^{22}$cm$^{-2}$        & &  &\\
\hline
050406(r) & $1.1\pm 0.4$ & $<0.09$      & 2.45$\pm$0.71 & 0.6 & \\
050406(d) & $-1.05 \pm 0.3$& $<0.06$ & 2.23$\pm$0.55 & 1.42 & IS\\
050502B(r) & $1.41\pm 0.03$ & 0.15$\pm$0.01     & 2.27$\pm$0.03 & 1.02 & \\
050502B(d) & -0.88 & 0.08$\pm$0.01      & 2.33$\pm$0.04 & 1.2 & IS\\
050713A(1r)& $0.67\pm 0.05$ &  $0.53\pm 0.09$  & 1.60 $\pm$ 0.15& 1.11& \\
050713A(1d)& -0.25 to -1 &  $0.62\pm 0.05$  & 2.60 $\pm$0.15 & 1.27 & RS case i\\
050730(1r)& $0.37\pm 0.12$ &  $1.6^{+2.5}_{-1.6}$  & 1.29 $\pm$ 0.16& 0.86 & \\
050730(1d)&$-0.41\pm0.06$&  $3.1^{+1.3}_{-1.1}$  & 1.82 $\pm$0.12 & 1.3 &\\
050730(2r)& $0.92\pm 0.10$ & $2.1^{+0.8}_{-0.7}$& 1.71 $\pm$ 0.08& 1.1 &\\
050730(2d)&$-0.67\pm0.16$ &  $0.9\pm0.5$  & 1.70 $\pm$0.12 & 1.1 & RS case iv\\
050730(3r)& $0.51\pm 0.12$ & $0.7^{+0.8}_{-0.7} $& 1.77 $\pm$ 0.12& 0.87 & \\
050730(3d)&$-0.27\pm 0.08$&  $1.0\pm0.6$  & 2.01 $\pm$0.1 & 0.81& \\
050822(r) & $0.45\pm 0.15$ & 0.24$\pm$0.05     & 2.8$\pm$0.2& 0.8 & \\
050822(d) & $-0.56\pm 0.08$ & 0.19$\pm$0.04      & 2.8$\pm$0.2 &0.9& IS\\
050904(r) & $0.25\pm 0.06$& 0.14$\pm$0.02     & 1.76$\pm$0.06 & 1.22 &\\
050904(d) & -0.64$\pm$ 0.05& 0.07$\pm$0.03      & 1.75$\pm$0.08 & 0.8 & RS case iv\\
060111A(r) & 0.000& 0.14$\pm$0.02     & 1.76$\pm$0.06 & 1.22 &\\
060111A(d)  & 0.48$\pm$ 0.03& 0.07$\pm$0.03      & 1.75$\pm$0.08 & 0.8 & RS\\
\hline
\end{tabular}
\end{table}

\begin{table}[ht!]
\caption{\bf Broken power law fits for the spectra. }
\begin{tabular}{l|cccccc}
\hline
\hline
GRB     &  N$_H$     &   $\Gamma_2$      &   E$_b$ &$\chi^2_{\rm R}$   \\
        & $10^{22}$cm$^{-2}$ &           &         &                   \\
\hline
050406(r)&na & na & na&na \\
050406(d)&na&na& na& na\\
050502B(r) & 0.12$\pm$ 0.02 &  2.25$\pm$ 0.04 & 0.71$\pm$ 0.08 &0.96  \\
050502B(d) & 0.04$\pm$ 0.02  & 2.29 $\pm$ 0.05 & 0.72$\pm$ 0.08 &1.09 \\
050713A(1r)&$0.50\pm 0.09 $  & $1.95^{+0.75}_{-0.35}$ & $3.75^{+1.5}_{-1.75} $& 1.18 \\
050713A(1d)& $0.62\pm 0.05$  & 2.67 $\pm$ 0.17& $1.8\pm 0.2$& 1.18 \\
050730(1r)& $1.6^{+2.5}_{-1.6} $  & 1.29 $\pm$ 0.15& $<2.2$& 0.89 \\
050730(1d)&$2.0^{+1.1}_{-0.8}$  & 1.85 $\pm$ 0.12& $1.1^{+0.7}_{-0.3}$& 1.28  \\
050730(2r)&na & na & na&na \\
050730(2d)&na&na& na& na\\
050730(3r)&na &na & na & na\\
050730(3d)&na&na&na& na \\
050822(r) & 0.2$\pm$ 0.1  & 2.8$\pm$ 0.3 & 0.7$\pm$ 0.3 &0.7\\
050822(d) & na & na & na & na \\
050904(r) & 0.13$\pm$ 0.04  & 1.73$\pm$ 0.07 & 0.7$\pm$ 0.6 &1.2 \\
050904(d) & 0.06$\pm$ 0.05  & 1.74$\pm$ 0.09 & 0.8$\pm$ 0.7 &0.8 \\
060111A(r) & 0.14$\pm$0.02     & 1.76$\pm$0.06 & 1.22 \\
060111A(d)  & 0.06$\pm$ 0.05  & 1.74$\pm$ 0.09 & 0.8$\pm$ 0.7 &0.8 \\
\hline
\end{tabular}
\end{table}

\section{Discussion}

It is by now apparent that flares in the light curves of GRBs decay
have  different characteristics. While it may prove difficult to explain
their complex phenomenology within the framework of a single model,
we can nevertheless constrain
some of the models by carefully exploiting the present data. 

We have presented a timing and spectral analysis of several flares 
detected by Swift. 
Since the main energy output of the GRB
engine rapidly decays with time \cite{ref:janiuk04}, the energy
injections are likely due either to  slower shells ejected during the
prompt phase of the GRB engine and catching up at later times (refreshed
shocks), or to
later energy production episodes in the central engine, due, e.g.,to
 disk fragments that accrete
at later times \cite{ref:perna06} (late internal shocks).
We have proposed a method to distinguish between these two different models
based on the estimate of the flare duration and on the comparison
between the spectral and temporal slopes.

We found that the flares of  GRB050713A, GRB050730 and GRB050904
can be explained by the refreshed shock model.
We saw that the flares in GRB050730
do not satisfy the relation implied by the curvature effect 
between the temporal and  spectral slope,
indicating that internal shocks may be less likely in this case.
However there are other bursts (like GRB 050406, GRB 050502B and GRB050822)
were late internal shocks
may interpret better  the flares characteristics. We plan 
to extend our analysis to a larger sample of Swift GRBs.

Flaring activity is also observed in the optical afterglow of some GRBs
(like GRB 050730 see Perri et al. 2006). 
Detection of optical rebrightenings in the light curve would 
also help in testing the models.
Simultaneous re-brightening events in the X--ray and 
optical bands, like the ones seen in GRB 050730 can be easily explained 
in the refreshed shock model (e.g. \cite{ref:Granot03}).
 
\bigskip

This research has been partially supported by A
SI grant I/R/039/04
and MIUR.


\begin{thebibliography}{0}

\bibitem{ref:laurs03}
\BY{Laursen~L.T. \atque Stanek~K.Z.} 
\IN{Astrophys. J.}{597}{2003}{L107}

\bibitem{ref:sari98}
\BY{Sari~R., Piran~T. \atque Narayan~R.}
\IN{Astrophys. J.}{497}{1998}{L17}

\bibitem{ref:bers02}
\BY{Bersier~D. et al.}
\IN{Astrophys. J.}{584}{2002}{L43}

\bibitem{ref:math02}
\BY{Matheson~T. et al. 2002}
\IN{Astrophys. J.}{582}{2002}{L5} 


\bibitem{ref:geh05}
\BY{Gehrels~N. et al}
\IN{Astrophys. J.}{621}{2005}{558}

\bibitem{ref:bur05}
\BY{Burrows~D. N. et al.}
\IN{Science}{309}{2005}{1833B}

\bibitem{ref:falc06}
\BY{Falcone~A.D. et al.}
\IN{Astrophys. J.}{641}{2006}{1010}


\bibitem{ref:nous06}
\BY{Nousek~J. A. et al.}
\IN{Astrophys. J.}{642}{2005}{389}


\bibitem{ref:obr06}
\BY{O'Brien~P. T. et al.}
\IN{Astrophys. J.}{647}{2006}{12130}


\bibitem{ref:rom06}
\BY{Romano~P. et al.} 
\IN{Astron. \atque Astrophys.}{456}{2006}{917}

\bibitem{ref:koba06}
\BY{Kobayashi~S. et al.}
astro-ph/0506157


\bibitem{ref:laz02}
\BY{Lazzati~D., et al.}
\IN{Astron. \atque Astrophys.}{396}{2002}{L5}

\bibitem {ref:zhang06}
\BY{Zhang~B. et al.}
\IN{Astrophys. J.}{642}{2006}{354}


\bibitem{ref:king06}
\BY{King~A. et al.}
\IN{Astrophys. J.}{630}{2005}{L113}


\bibitem{ref:mac05}
\BY{MacFadyen~A. I., Ramirez-Ruiz~E. \atque Zhang~W.}
astro-ph/0510192


\bibitem{ref:perna06}
\BY{Perna~R. Armitage~P. J. \atque Zhang~B.}
 \IN{Astrophys. J.}{636}{2006}{L29}


\bibitem{ref:gao06}
\BY{Gao~W. H. \atque Fan~Y.Z.}
astro-ph/0512646 submitted to ApJ


\bibitem{ref:proga06}
\BY{Proga~D. \atque Zhang~B.}
\IN{MNRAS}{370}{L61}

\bibitem{ref:kum00}
\BY{Kumar~P. \atque Piran~T.}
\IN{Astrophys. J.}{532}{2000}{286}

\bibitem{ref:guetta06}
\BY{Guetta~D. et al.}
\IN{Astron. \atque Astrophys.}{2006}{in}{press}

\bibitem{ref:perri06}
\BY{Perri~M. et al.}
\TITLE{submitted to Astron. and Astrophys.}

\bibitem{ref:guetta06b}
\BY{Guetta~D. et al.}
\TITLE{in preparation}

\bibitem{ref:capa05}
\BY{Capalb~M. et al.}
\TITLE{The SWIFT XRT Data Reduction Guide}
http://heasarc.gsfc.nasa.gov/docs/swift/analysis/xrt\_swguide\_v1\_2.pdf

\bibitem{ref:liang06}
\BY{Liang~E.W. et al.}
\IN{Astrophys. J.}{646}{2006}{L351}

\bibitem{ref:kumar00b}
\BY{Kumar~P. \atque Panaitescu}
\IN{Astrophys. J.}{541}{2000}{L51}


\bibitem{ref:janiuk04}
\BY{Janiuk~A., Perna~R., Di Matteo~T. \atque Czerny~B.}
\IN{MNRAS}{355}{2004}{950}


\bibitem{ref:Granot03}
\BY{Granot~J., Nakar~E. \atque Piran~T.}
\IN{Nature}{426}{2003}{138}

\bibitem{ref:chinca06}
\BY{Chincarini~G. et al}
in preparation


\end{thebibliography}
\end{document}